\documentclass[12pt,]{article}
\usepackage{epsf,latexsym,rotate}
\usepackage[ps,arc,frame]{xy}
\usepackage{amsfonts,amssymb} 
\epsfverbosetrue
\usepackage[ps,arc,frame]{xy}
\usepackage{graphicx}
\usepackage{slashed}

\newcommand{\lapprox}{%
\mathrel{%
\setbox0=\hbox{$<$}
\raise0.6ex\copy0\kern-\wd0
\lower0.65ex\hbox{$\sim$}
}}
\newcommand{\gapprox}{%
\mathrel{%
\setbox0=\hbox{$>$}
\raise0.6ex\copy0\kern-\wd0
\lower0.65ex\hbox{$\sim$}
}}

\newcommand{\double}[1]{\mathbb{#1}}

\newcommand{\zz}{\double{Z}}

\newcommand{\ul}[1]{\underline{#1}}

\newcommand{\op}{\oplus}

\newcommand{\bb}{\begin{eqnarray}}
\newcommand{\ee}{\end{eqnarray}}
\newcommand{\eee}{\nonumber\end{eqnarray}}
\newcommand{\qq}{\quad}

\begin{document}

\font\twelve=cmbx10 at 13pt
\font\eightrm=cmr8

\thispagestyle{empty}

\begin{center}

CENTRE DE PHYSIQUE TH\'EORIQUE $^1$ \\ CNRS--Luminy, Case
907\\ 13288 Marseille Cedex 9\\ FRANCE\\

\vspace{2cm}

{\Large\textbf{Higgs-mass predictions}} \\

\vspace{1.5cm}

{\large  Thomas
Sch\"ucker $^2$}

\vspace{2cm}

{\large\textbf{Abstract}}

\vspace{.44cm}

A compilation of Higgs-mass predictions is proposed.
\end{center}

\vskip 5 truecm

\noindent
PCAC-06: 14.80.Bn  Standard-model Higgs boson

\vskip 2truecm

\noindent 0708.3344\\

\vspace{1.5cm}
\noindent $^1$ Unit\'e Mixte de Recherche  (UMR 6207)
du CNRS  et des Universit\'es Aix--Marseille 1 et 2 et  Sud
Toulon--Var, Laboratoire affili\'e \`a la FRUMAM (FR 2291)\\
$^2$ also at Universit\'e Aix--Marseille 1,
thomas.schucker@gmail.com \\

\section{Introduction}

Many physicists hope that the electro-weak Higgs scalar will be observed soon at the LHC. The literature contains a plethora of predictions or upper limits of the Higgs mass based on many different ideas, models and calculational techniques. 
Privileged among them is the value ${ m_H=150\  \pm\  36\ {\rm GeV}}$ currently given by
the LEP Electroweak Working Group, because it only relies on precision electro-weak data,
non-observation of the Higgs today and the minimal hypothesis that the standard model is correct as it stands.

 A compilation of all other predictions is attempted here. Some models make additional predictions or postdictions, that are indicated. The point in time separating pre- and postdiction is taken as the time of publication of the model. In this task it is unavoidable to miss references and I acknowledge the feed back to the successive versions from colleagues.

The predictions are organised in increasing order of the central value of the predicted mass interval. In a second section the upper limits are presented in increasing order as well. A third section contains two lower limits.  Older predictions and limits incompatible with today's experimental lower limit of 114 GeV are not recorded here. As another example not covered I should mention the supersymmetric model by 
 Derm\'{i}\v{s}ek \& Gunion (2005)  with a Higgs mass of 100 GeV. Because of exotic decay channels this model is still compatible with LEP data.  Also not recorded are predictions that come with postdictions contradicting present experimental numbers.

The references are in alphabetical order of the first author's last name with first name and date  as secondary criteria.

\section{Predictions}

$\bullet  \qq m_H=109\  \pm$ 12 GeV\\
{\bf Authors:} O.~Buchm\"uller {\it et al.} (2007)\\
{\bf Idea:} constrained minimal supersymmetric standard model combined with electro-weak precision data, flavor physics and abundance of cold dark matter\\
{\bf Techniques:} multi-parameter fit, renormalisation group equation. The top mass is taken to be $m_t=170.9\pm 1.8$ GeV. \\
{\bf Other predictions:} many supersymmetric particles\\[2mm]
$\bullet  \qq m_H=115.3\  \pm$ 0.1 GeV\\
{\bf Author:} Sch\"ucking (2007)\\
{\bf Idea:} interpretation of the $SU(2)\times U(1)$ group of the electro-weak forces as symmetry group of the Eguchi-Hanson metric\\
{\bf Techniques:} differential geometry and quaternions \\[2mm]
$\bullet  \qq m_H=115.4\  \pm$ 0.9 GeV\\
{\bf Author:} Popovic (2010)\\
{\bf Idea:} top quark as bound state of 3 prequarks, Higgs of 2 prequarks\\
{\bf Techniques:} arithmetic \\[2mm]
$\bullet  \qq m_H=115.9\  \pm$ 2 GeV\\
{\bf Authors:} Cassel \& Ghilencea
 (2011)\\
{\bf Idea:} supersymmetry\\
{\bf Techniques:} constrained minimal supersymmetric extension of the standard model plus consistency of the lightest supersymmetric particle as dark matter with the WMAP data\\
{\bf Other predictions:} many supersymmetric particles\\[2mm]
$\bullet  \qq m_H=117\  \pm$ 4 GeV \\
{\bf Authors:} Gogoladze, Okada \& Shafi (2007)\\
{\bf Idea:} Higgs boson as zero mode of gauge boson along a fifth compactified dimension\\
{\bf Techniques:} a boundary condition on the Higgs self-coupling at compactification scale $\Lambda$ and renormalisation group flow up to energies of  $\Lambda\sim 10^8$ GeV \\[2mm]
$\bullet  \qq m_H=117\  \pm$ 12 GeV\\
{\bf Authors:} Kane, Kumar, Lu \& Zheng
 (2011)\\
{\bf Idea:} compactified string/M theories\\
{\bf Techniques:}  minimal supersymmetric extension of the standard model \\
{\bf Other predictions:} many supersymmetric particles\\[2mm]
$\bullet  \qq m_H=118$ \\
{\bf Authors:} Arbuzov, Barbashov, Pervushin, Shuvalov \& Zakharov (2008)\\
{\bf Idea:} Three peaks of the cosmic microwave background are explained by the decay of primordial Higgs-, $W$- and $Z$-bosons into photons. \\
{\bf Techniques:} conformal cosmology\\
{\bf Other prediction:} $m_H=216$ \\[2mm]
$\bullet  \qq m_H=120\  \pm$ 6 GeV\\
{\bf Authors:} Ellis, Nanopoulos, Olive \& Santoso (2005)\\
{\bf Idea:} supersymmetry\\
{\bf Techniques:} minimal supersymmetric extension of the standard model with universal soft supersymmetry-breaking masses\\
{\bf Other predictions:} many supersymmetric particles\\[2mm]
$\bullet  \qq m_H=121\  \pm$ 6 GeV\\
{\bf Authors:} Feldstein, Hall \& Watari (2006)\\
{\bf Idea:} superstring inspired landscape of vacua and some probability density for the parameters of the Higgs potential\\
{\bf Techniques:} renormalisation group flow up to energies of  $\Lambda\sim 10^{19}$ GeV\\
{\bf Postdiction:}  $m_t=176\  \pm 2$ GeV
\\[2mm]
$\bullet  \qq m_H=121.25\  \pm$ 2.25 GeV\\
{\bf Authors:} Li, Maxin, Nanopoulos \& Walker (2011)\\
{\bf Idea:} supersymmetry\\
{\bf Techniques:} F-theory no scale supergravity and $SU(5)$\\
{\bf Other predictions:} many supersymmetric particles\\[2mm]
$\bullet  \qq m_H=121.8\  \pm$ 11 GeV\\
{\bf Authors:} Froggatt \& Nielsen (1995)\\
{\bf Idea:} two approximately degenerate vacua, one in which we live, the other of Planck energy\\
{\bf Techniques:} renormalisation group equations\\
{\bf Postdiction:}  $m_t=173\  \pm 4$ GeV\\[2mm]
$\bullet  \qq m_H=122\  \pm$ 10 GeV\\
{\bf Authors:} Djouadi, Heinemeyer, Mondragon \& Zoupanos (2004)\\
{\bf Idea:}  a supersymmetric version of $SU(5)$\\
{\bf Techniques:} renormalisation group flow up to energies of  $\Lambda\sim 10^{16}$ GeV\\
{\bf Other predictions:} many supersymmetric particles\\
{\bf Postdictions:} $m_t=174-183$ GeV\\[2mm]
$\bullet  \qq m_H=122.8$ GeV\\
{\bf Author:} Bogan (2009)\\
{\bf Idea:} simple relations between cosmological constant, GUT scale and  masses of the electron, inflaton, and Higgs\\
{\bf Techniques:} geometric mean
\\[2mm]
$\bullet  \qq m_H=123$ GeV\\
{\bf Author:} Stech (2010)\\
{\bf Idea:}  $SO(10)$ or $E_6$ grand unification plus a $SO(3)$ flavour symmetry
\\
{\bf Techniques:} group representations\\[2mm]
$\bullet  \qq m_H=123.5\  \pm$ 5.5 GeV\\
{\bf Authors:} Heinemeyer, M.~Mondragon \& G.~Zoupanos (2007)\\
{\bf Idea:}  a supersymmetric Grand Unified Theory that can be made all-loop finite\\
{\bf Techniques:} renormalisation group flow with $m_t=170.9$ GeV\\
{\bf Other predictions:} supersymmetric particles\\
$\bullet  \qq m_H=124\  \pm$ 21 GeV\\
{\bf Authors:} Barger, Deshpande, Jiang, Langacker \& Li (2007)\\
{\bf Idea:} supersymmetry broken at $10^{5}-10^{16}$ GeV and gauge coupling unification at $\Lambda\sim 10^{16}-10^{17}$ GeV \\
{\bf Techniques:} renormalisation group flow up to energies of  $\Lambda$ \\
{\bf Other predictions:} new vectorlike fermions with masses in the 200 - 1000 GeV range
\\[2mm]
$\bullet  \qq m_H=124\  \pm$ 10 GeV \\{\bf Authors:} Arbuzov, Glinka, Lednicky \& Pervushin (2007), version 6\\
{\bf Idea:} condensates, conformal cosmology \\
{\bf Techniques:} Coleman-Weinberg potential \\
{\bf Other predictions:} $m_H=275\pm 25$ GeV, version 1\\[2mm]
$\bullet  \qq m_H=124.2\  \pm$ 13.2 GeV\\
{\bf Authors:} Codoban, Jurcisin \& Kazakov (1999)\\
{\bf Idea:} supersymmetry\\
{\bf Techniques:} minimal supersymmetric extension of the standard model with non-universal soft supersymmetry-breaking masses\\
{\bf Other predictions:} many supersymmetric particles\\[2mm]
\vfil\eject
$\bullet  \qq m_H=125\  \pm$ 5 GeV\\
{\bf Authors:} Kahana \& Kahana (1993)\\
{\bf Idea:} dynamical symmetry breaking and the Higgs as a deeply bound state of two top quarks\\
{\bf Techniques:} Nambu Jona-Lasinio theory\\
{\bf Other predictions:} $m_t=175\ \pm\ 5$ GeV. Note that the top was discovered in 1995.
\\[2mm]
$\bullet  \qq m_H=125\  \pm$ 4 GeV\\
{\bf Authors:} Gogoladze, Okada \& Shafi (2007)\\
{\bf Idea:} Higgs boson as zero mode of gauge boson along a fifth compactified dimension\\
{\bf Techniques:} a boundary condition on the Higgs self-coupling at compactification scale $\Lambda$ and renormalisation group flow up to energies of  $\Lambda\sim  10^{13}-10^{14}$ GeV. \\[2mm]
$\bullet  \qq m_H=126.3\  \pm$ 2.2 GeV\\
{\bf Authors:} Shaposhnikov \& C.~Wetterich (2009)\\
{\bf Idea:} Assume that gravity is asymptotically safe, that there are no intermediate energy scales between the Fermi and Planck scales, that the gravity induced anomalous dimension of the Higgs selfcoupling is positive.  \\
{\bf Techniques:}  renormalisation group flow with $m_t=171.2 $ GeV \\[2mm]
$\bullet  \qq m_H=127.5\  \pm$ 7.5 GeV\\
{\bf Authors:} Chankowski, Falkowski, Pokorski \& Wagner (2004)\\
{\bf Idea:} supersymmetry and the Higgs as a pseudo-Goldstone boson of some extra global symmetry\\
{\bf Techniques:} smaller fine-tuning than in the minimal supersymmetric extension of the standard model.
The computation of the Higgs mass depends on the top mass taken to be $m_t=178 \pm 4.3$ GeV. \\
{\bf Other predictions:} many supersymmetric particles and an additional $Z$-boson with a mass of 3 TeV\\[2mm]
 $\bullet  \qq m_H=129.6$  GeV\\
{\bf Authors:}  X. Calmet \& H. Fritzsch (2001)\\
{\bf Idea:} confining $SU(2)$ and a `complementarity principle'\\
{\bf Techniques:} 1-loop corrections \\[2mm]
$\bullet  \qq m_H=130\  \pm$ 6 GeV\\
{\bf Authors:} Dae Sung Hwang, Chang-Yeong Lee \&  Ne'eman (1996)\\
{\bf Idea:} embedding of the electro-weak Lie algebra $su(2)\op u(1)$ in the superalgebra $su(2|1)$\\
{\bf Techniques:} renormalisation group flow \\
{\bf Postdictions:} $\sin^2\theta_w=0.229\pm 0.005$
\\[2mm]
$\bullet  \qq m_H=130\  \pm$ 10 GeV \\
{\bf Authors:} Nakayama \& F.~Takahashi (2011)
\\
{\bf Idea:} Identify the Higgs with the inflaton and use correlation between Higgs mass and a spectral index of density perturbations of 0.95 -- 0.96. \\
{\bf Techniques:} PeV scale supersymmetry breaking \\
{\bf Other predictions:} many supersymmetric particles\\[2mm]
$\bullet  \qq m_H=131\  \pm$ 10 GeV \\
{\bf Authors:} Gogoladze, Li, Senoguz \& Shafi (2006)\\
{\bf Idea:} 7 dimensional orbifold with $SU(7)$ grand unification and split supersymmetry \\
{\bf Techniques:} a boundary condition on the Higgs self-coupling at unification scale $\Lambda$ and renormalisation group flow up to energies of  $\Lambda\sim 10^{16}$ GeV \\
{\bf Other predictions:} $m_H=146\  \pm$ 8 GeV\\[2mm]
$\bullet  \qq m_H=134\  \pm$ 9 GeV \\
{\bf Authors:} Ni, Lou, Lu \& Yang (1998)\\
{\bf Idea:} a Gau\ss ian effective potential \\
{\bf Techniques:} renormalisation group flow up to energies of  $\Lambda\sim 10^{15}$ GeV \\[2mm]
$\bullet  \qq m_H=135\  \pm$ 6 GeV \\
{\bf Authors:} Gogoladze, Li \& Shafi (2006)\\
{\bf Idea:} 7 dimensional  $N=1$ supersymmetric orbifold with $SU(7)$ grand unification \\
{\bf Techniques:} a boundary condition on the Higgs self-coupling at unification scale $\Lambda$ and renormalisation group flow up to energies of  $\Lambda\sim 10^{16}$ GeV \\
{\bf Other predictions:} $m_H=144\  \pm$ 4 GeV\\[2mm]
 $\bullet  \qq m_H=135\  \pm$ 15 GeV \\
{\bf Authors:} Arkani-Hamed \& Dimopoulos (2004)\\
{\bf Idea:} split supersymmetry \\
{\bf Techniques:} fine tuning and renormalisation group flow up to energies of  $ 10^{16}$ GeV \\[2mm]
 $\bullet  \qq m_H=137\  \pm$ 23 GeV \\
{\bf Authors:}  Medina,  Shah \& Wagner (2007)\\
{\bf Idea:} a warped fifth dimension and an extension of the electro-weak gauge symmetry to $SO(5)\times U(1)$ in the bulk, broken at the boundaries \\
{\bf Techniques:} Coleman-Weinberg potential \\[2mm]
$\bullet  \qq m_H=140\  \pm$ 10 GeV \\
{\bf Authors:} Babu, Gogoladze, Rehman \& Shafi (2008)
\\
{\bf Idea:} minimal supersymmetric extension of the standard model plus complete vectorlike multiplets of grand unified groups\\
{\bf Techniques:} some fine tuning and renormalisation group flow up to energies of  $ 10^{16}$ GeV. The top mass is taken to be $m_t=172.6\pm1.4 $ GeV.  \\
{\bf Other predictions:} many supersymmetric particles\\[2mm]
$\bullet  \qq m_H=141\  \pm$ 2 GeV \\
{\bf Authors:} Hall \& Y.~Nomura (2009)
\\
{\bf Idea:} minimal supersymmetric extension of the standard model plus supersymmetry breaking at very high scale,  motivated from a multiverse\\
{\bf Techniques:} huge fine tuning and 2-loop corrections with a top mass of 173.1 GeV \\
{\bf Other predictions:} no supersymmetric particles\\[2mm]
$\bullet  \qq m_H=143\  \pm$ 37 GeV\\
{\bf Authors:} Cabibbo, Maiani, Parisi \& Petronzio (1979)\\
{\bf Idea:} the big desert: no new particles besides the Higgs and validity of perturbative quantum field theory up to the Planck scale \\
{\bf Techniques:} renormalisation group flow up to energies of  $ 10^{19}$ GeV. The computation of the Higgs mass depends on the top mass taken here to be $m_t=171.5\pm 2$ GeV.
\\[2mm]
$\bullet  \qq m_H=143.2\  \pm$ 28.8 GeV\\
{\bf Authors:} Gogoladze, Okada \& Shafi (2007b)\\
{\bf Idea:} 2 extra dimensions compactified on an orbifold\\
{\bf Techniques:} a boundary condition on the Higgs self-coupling at compactification scale $\Lambda$ and renormalisation group flow up to energies of  $\Lambda\sim  10^{19}$ GeV. 
\\[2mm]
$\bullet  \qq m_H=143.4\  \pm$ 1.3 GeV\\
{\bf Author:} Popovic (2010)\\
{\bf Idea:} radiatively generated Higgs mass\\
{\bf Techniques:} cancellation of certain leading divergences 
\\[2mm]
$\bullet  \qq m_H=144\  \pm$ 4 GeV \\
{\bf Authors:} Gogoladze, Li \& Shafi (2006)\\
{\bf Idea:} 7 dimensional  $N=1$ supersymmetric orbifold with $SU(7)$ grand unification \\
{\bf Techniques:} a boundary condition on the Higgs self-coupling at unification scale $\Lambda$ and renormalisation group flow up to energies of  $\Lambda\sim 10^{16}$ GeV. \\
{\bf Other predictions:} $m_H=135\  \pm$ 6 GeV\\[2mm]
$\bullet  \qq m_H=145\  \pm$ 7 GeV \\
{\bf Author:} Liu (2005)\\
{\bf Idea:} supersymmetry broken at $10^{11}$ GeV and a $\zz_3$ symmetry among generations \\
{\bf Techniques:} radiative corrections \\[2mm]
$\bullet  \qq m_H=146\  \pm$ 8 GeV \\
{\bf Authors:} Gogoladze, Li, Senoguz \& Shafi (2006)\\
{\bf Idea:} 7 dimensional orbifold with $SU(7)$ grand unification and split supersymmetry \\
{\bf Techniques:} a boundary condition on the Higgs self-coupling at unification scale $\Lambda$ and renormalisation group flow up to energies of  $\Lambda\sim 10^{16}$ GeV. \\
{\bf Other predictions:} $m_H=131\  \pm$ 10 GeV\\[2mm]
$\bullet  \qq m_H=146\  \pm$ 19 GeV \\
{\bf Authors:} Barger, Jiang, Langacker \& Li (2005)\\
{\bf Idea:} supersymmetry broken at high scale and gauge coupling unification at $\Lambda\sim 10^{16}-10^{17}$ GeV \\
{\bf Techniques:} renormalisation group flow up to energies of  $\Lambda$ 
\\[2mm]
$\bullet  \qq m_H=148.1\  \pm$ 10.7 GeV\\
{\bf Author:} Popa (2009)\\
{\bf Idea:}  Let the Higgs be the inflaton by adding a strong, non-minimal coupling $\varphi ^2R$ of the scalar to gravity. 
\\
{\bf Techniques:} effective action with $m_t=171.3$ GeV and its confrontation with the observed spectral index and tensor-to-scalar ratio of the Cosmic Microwave Background \\[2mm]
 $\bullet  \qq \ul{ m_H=150\  \pm\  36\ {\rm GeV}}$\\[1mm]
{\bf Authors:} The LEP Electroweak Working Group \\
{\bf Idea:} non-observation of the Higgs and quantum corrections by Higgs loops to precision electro-weak data\\
{\bf Techniques:} experiment and quantum field theory 
\\[2mm]
$\bullet  \qq m_H=150\  \pm$ 10 GeV \\
{\bf Authors:} Barger, Chiang, Jiang \& Li (2004)\\
{\bf Idea:} supersymmetry broken at $10^{11}$ GeV and Peccei-Quinn symmetry \\
{\bf Techniques:} radiative corrections \\[2mm]
$\bullet  \qq m_H=150\  \pm$ 20 GeV \\
{\bf Authors:} Arvanitaki, Davis, Graham \& Wacker (2004)\\
{\bf Idea:} split supersymmetry \\
{\bf Techniques:} fine tuning and renormalisation group flow up to energies of  $ 10^{16}$ GeV \\[2mm]
$\bullet  \qq m_H=150\  \pm$ 50 GeV\\
{\bf Authors:} Bai, Fan \& Han (2007)\\
{\bf Idea:} supersymmetry and a long-lived metastable supersymmetry breaking vacuum\\
{\bf Techniques:} little Higgs mechanism, 1-loop corrections\\
{\bf Other predictions:} many supersymmetric particles plus new gauge bosons and electro-weak triplets at 1 TeV  \\[2mm]
$\bullet  \qq m_H=150\  \pm$ 24 GeV\\
{\bf Authors:} Shaposhnikov \& Wetterich (2009)\\
{\bf Idea:} Assume that gravity is asymptotically safe, that there are no intermediate energy scales between the Fermi and Planck scales.  \\
{\bf Techniques:}  renormalisation group flow with $m_t=171.2 $ GeV \\[2mm]
$\bullet  \qq m_H=150$  GeV\\
{\bf Authors:} Chiang \& Nomura (2010)\\
{\bf Idea:} $E_6$ unification in six dimensions and $S^2/\zz_2$ orbifold compactification \\
{\bf Techniques:}  tree level masses from Kaluza-Klein model\\[2mm]
$\bullet  \qq m_H=153\  \pm$ 3 GeV\\
{\bf Author:} Okumura (1997)\\
{\bf Idea:} a vague variant of the Connes-Lott model\\
{\bf Techniques:} 2-loop renormalisation group flow up to energies of  $ 10^{13}$ GeV. The computation of the Higgs mass depends on the top mass taken here to be $m_t=171.5\pm 2$ GeV.
\\[2mm]
$\bullet  \qq m_H=154\  \pm$ 6 GeV \\
{\bf Authors:} Ananthanarayan \& Pasupathy (2001)\\
{\bf Idea:} weak dependence of the ratio between Higgs self-coupling and top Yukawa coupling squared on renormalisation scale \\
{\bf Techniques:} 1- and 2-loop corrections \\[2mm]
$\bullet  \qq m_H=154\  \pm$ 37 GeV\\
{\bf Authors:} Gogoladze, He \& Shafi (2010)
\\
{\bf Idea:} Vectorlike isospin doublets of quarks with masses of several 100 GeV are added to the standard model to achieve gauge unification at $3\cdot 10^{16}$ GeV. Assuming the validity of perturbative quantum field theory up to this energy constrains the Higgs-mass as in Cabibbo, Maiani, Parisi \& Petronzio (1979).  \\
{\bf Techniques:} renormalisation group flow\\
{\bf Other predictions:} these new vectorlike quarks
\\[2mm]
$\bullet  \qq m_H=154.4\  \pm$ 0.5 GeV\\
{\bf Author:} Beck (2001)\\
{\bf Idea:} `chaotic strings'
 describing the dynamics of vacuum fluctuations underlying
dark energy \\
{\bf Techniques:} stochastic quantization\\
{\bf Postdictions:} all fermion and gauge bosons masses, all gauge couplings
\\[2mm]
$\bullet  \qq m_H=155\  \pm$ 8 GeV\\
{\bf Authors:} Schrempp \& Schrempp (1993)\\
{\bf Idea:} A strongly infrared attractive line in the $m_t-m_H$ plane is found.\\
{\bf Techniques:} 1-loop renormalization group equations
\\[2mm]
$\bullet  \qq m_H=160\  \pm$ 8 GeV\\
{\bf Authors:} Roepstorff \& Vehns (2000)\\
{\bf Idea:} combining gauge and Yukawa interactions in one generalised Dirac operator \\
{\bf Techniques:} superconnections\\
{\bf Postdictions:} $m_t=160\  \pm$ 8 GeV
\\[2mm]
$\bullet  \qq m_H=160\  \pm$ 20 GeV \\
{\bf Authors:} Langacker, Paz, Wang \& Yavin (2007)\\
{\bf Idea:} an extension of the minimal supersymmetric extension of the standard model plus a hidden sector plus a $Z'$ mediating supersymmetry breaking by couplings to the hidden sector
 \\
{\bf Techniques:} 2-loop corrections up to energies of  $10^7-10^{11}$ GeV\\
{\bf Other predictions:} many supersymmetric particles
\\[2mm]
$\bullet  \qq m_H=160\  \pm$ 24.5 GeV \\
{\bf Authors:} Barvinsky, Kamenshchik, Kiefer, Starobinsky \& Steinwachs (2009)\\
{\bf Idea:}  Let the Higgs be the inflaton by adding a strong, non-minimal coupling $\varphi ^2R$ of the scalar to gravity. 
\\
{\bf Techniques:} effective action to 1-loop with $m_t=171$ GeV and its confrontation with the observed spectral index and tensor-to-scalar ratio of the Cosmic Microwave Background  \\[2mm]
$\bullet  \qq m_H=160\  \pm$ 30 GeV \\
{\bf Author:} Bezrukov (2008)\\
{\bf Idea:}  Let the Higgs be the inflaton by adding a strong, non-minimal coupling $\varphi ^2R$ of the scalar to gravity. 
\\
{\bf Techniques:} effective action  and its confrontation with the observed spectral index and tensor-to-scalar ratio of the Cosmic Microwave Background  \\[2mm]
$\bullet  \qq m_H=160\  \pm$ 34 GeV \\
{\bf Authors:} Bezrukov \& Shaposhnikov (2009)\\
{\bf Idea:}  Let the Higgs be the inflaton by adding a strong, non-minimal coupling $\varphi ^2R$ of the scalar to gravity. 
\\
{\bf Techniques:} effective action to 2-loop with $m_t=171.2$ GeV and its confrontation with the observed spectral index and tensor-to-scalar ratio of the Cosmic Microwave Background  \\[2mm]
$\bullet  \qq m_H=160.7\  \pm$ 24 GeV \\
{\bf Authors:} Bezrukov, Magnin \& Shaposhnikov (2008)\\
{\bf Idea:}  Let the Higgs be the inflaton by adding a strong, non-minimal coupling $\varphi ^2R$ of the scalar to gravity. 
\\
{\bf Techniques:} effective action to 1-loop with $m_t=171$ GeV and its confrontation with the observed spectral index and tensor-to-scalar ratio of the Cosmic Microwave Background  \\[2mm]
$\bullet  \qq m_H= 160.9\ \pm $ 0.1 GeV\\
{\bf Author:} Ne'eman (1986)\\
{\bf Idea:} embedding of the electro-weak Lie algebra $su(2)\op u(1)$ in the superalgebra $su(2|1)$\\
{\bf Techniques:} classical field theory \\
{\bf Postdictions:} $\sin^2\theta_w=1/4$
\\[2mm]
$\bullet  \qq m_H= 161.8033989 $ GeV\\
{\bf Author:} El Naschie (2005)\\
{\bf Idea:} E-infinity theory\\
{\bf Techniques:} ? 
\\[2mm]
$\bullet  \qq m_H=165\  \pm$ 5 GeV \\
{\bf Authors} El Naschie (2005b) 
\\
{\bf Idea:} minimal supersymmetric extension of the standard model plus 'Pauli's principle of bidivision
and symmetry reduction'\\
{\bf Techniques:} no quantum corrections
\\[2mm]
$\bullet  \qq m_H=170\  \pm$ 10 GeV\\
{\bf Authors:} Chamseddine \& Connes (1996)\\
{\bf Idea:} derivation of the standard model from gravity by generalising Riemannian to noncommutative geometry\\
{\bf Techniques:} operator algebras, heat-kernel expansion and renormalisation group flow up to energies of  $\Lambda\sim 10^{13}-10^{17}$ GeV. The computation of the Higgs mass depends on the top mass taken to be $m_t=170.9\pm 2.5$ GeV.\\
{\bf Other predictions:} conceptual uncertainty in proper time measurements of $\Delta \tau\sim\hbar
/\Lambda=10^{-41}-10^{-37}$ s\\
{\bf Postdictions:} $m_W^2/(\cos^2\theta_w\,m_z^2)=1$, gluons must be massless and must have pure vector-couplings, $m_t<186.3$ GeV.
\\[2mm]
$\bullet  \qq m_H=177.5\  \pm$ 7.5 GeV\\
{\bf Authors:} Antusch, Kersten, Lindner \& Ratz (2002)\\
{\bf Idea:} the Higgs as a
composite particle from neutrino condensation\\
{\bf Techniques:} seesaw mechanism, gap equation, renormalisation group flow up to the condensation scale  $\Lambda = 10^{16}$ GeV
\\[2mm]
$\bullet  \qq m_H=182\  \pm$ 4 GeV\\
{\bf Author:} Namsrai (1996)\\
{\bf Idea:} Higgs mass from space-time curvature\\ 
{\bf Techniques:} general relativity and solitons
\\[2mm]
$\bullet  \qq m_H=185\  \pm$ 5 GeV\\
{\bf Author:} Schrempp \& Schrempp (1986)\\
{\bf Idea:} A largely unspecified strong interaction is assumed to soften the elastic scattering of longitudinally polarised $W$ bosons.\\
{\bf Techniques:} a superconvergence sum rule
\\[2mm]
$\bullet  \qq m_H=185.7\  \pm$ 0.1 GeV\\
{\bf Author:} Trostel (1987)\\
{\bf Idea:} a geometrisation of the Yukawa couplings\\
{\bf Techniques:} spinor connections
\\[2mm]
$\bullet  \qq m_H=186\  \pm$ 8 GeV\\
{\bf Authors:} Tolksdorf \& Thumst\"adter (2006)\\
{\bf Idea:} differential geometric unification of general relativity and the standard model\\
{\bf Techniques:} generalised Dirac operators, heat kernel expansion and renormalisation group flow up to energies of  $\Lambda\sim 10^{10}$ GeV. The computation of the Higgs mass depends on the top mass taken to be $m_t=174\pm 3$ GeV.
\\[2mm]
$\bullet  \qq m_H=194\  \pm$ 80 GeV\\
{\bf Authors:} Garc\'ia-Bellido, Figueroa \& Rubio (2008)\\
{\bf Idea:}  Let the Higgs be the inflaton by adding a strong, non-minimal coupling $\varphi ^2R$ of the scalar to gravity. 
\\
{\bf Techniques:} effective action and its confrontation with the observed spectral index and tensor-to-scalar ratio of the Cosmic Microwave Background, lower limit from non-observation of the Higgs at LEP  \\[2mm]
$\bullet  \qq m_H=196$  GeV\\
{\bf Authors:} Chiang, Nomura \&Sato (2011)\\
{\bf Idea:} $SO(12)$ unification in six dimensions and $S^2/\zz_2$ orbifold compactification \\
{\bf Techniques:}  tree level masses from Kaluza-Klein model\\[2mm]
$\bullet  \qq m_H=197.2\  \pm$ 124.8 GeV\\
{\bf Authors:} Froggatt, Laperashvili, Nevzorov, Nielsen \& Sher (2006)\\
{\bf Idea:} non-supersymmetric extension of the standard model with two Higgs doublets and the multiple point principle\\
{\bf Techniques:} renormalisation group flow up to energies of  $\Lambda\sim 10^{4}-10^{19}$ GeV\\
{\bf Other predictions:} additional neutral and charged scalars with masses larger than 202.4 GeV
 \\[2mm]
 $\bullet  \qq m_H=200\  \pm$ 20 GeV\\
{\bf Authors:} Froggatt, Nevzorov, Nielsen \& Thompson (2008)\\
{\bf Idea:} non-supersymmetric extension of the standard model with two Higgs doublets and the multiple point principle\\
{\bf Techniques:} renormalisation group flow up to energies of  $\Lambda\sim 10^{5}$ GeV. The computation of the Higgs mass depends on the top mass taken to be $m_t=171.4\pm 2.1$ GeV.\\
{\bf Other predictions:} enhanced top-Higgs coupling
 \\[2mm]
 $\bullet  \qq m_H=200\  \pm$ 50 GeV\\
{\bf Author:} Cveti\v{c} (1995)\\
{\bf Idea:} It is supposed that the 1-loop contributions of the scalar self-interactions to the effective potential are distinctly less than those of the Yukawa couplings of the top.\\
{\bf Techniques:} 1-loop corrections with cut-off at  $10^{3}$ GeV. The computation of the Higgs mass depends on the top mass taken to be $m_t=180$ GeV.
\\[2mm]
$\bullet  \qq m_H=203\  \pm$ 2.2 GeV\\
{\bf Author:} Squellari \& Stephan (2007)\\
{\bf Idea:} extension of Chamseddine and Connes' spectral action to include three vectorlike isospin doublets\\
{\bf Techniques:} operator algebras, heat-kernel expansion and renormalisation group flow up to  $\Lambda = 3-5\cdot 10^{7}$ GeV. The computation of the Higgs mass depends on the top mass taken to be $m_t=170.9\pm 2.6$ GeV.\\
{\bf Other predictions:} six new leptons with masses of $10-550$ TeV, conceptual uncertainty in proper time measurements of $\Delta \tau\sim\hbar
/\Lambda \sim  10^{-32}$ s\\
{\bf Postdictions:} $m_W^2/(\cos^2\theta_w\,m_z^2)=1$, gluons must be massless and must have pure vector-couplings.
\\[2mm]
$\bullet  \qq m_H=210\  \pm$ 10 GeV\\
{\bf Authors:} Andrianov \& Romanenko (1995)\\
{\bf Idea:} modified Veltman condition and fixed point in running of Yukawa coupling\\
{\bf Techniques:}  renormalisation group flow up to energies of  $10^{16}$ GeV\\
{\bf Postdictions:} $m_t= 175\pm 5$ GeV
\\[2mm]
$\bullet  \qq m_H=216$ \\
{\bf Authors:} Arbuzov, Barbashov, Pervushin, Shuvalov \& Zakharov (2008)\\
{\bf Idea:} Three peaks of the cosmic microwave background are explained by the decay of primordial Higgs-, $W$- and $Z$-bosons into photons. \\
{\bf Techniques:} conformal cosmology\\
{\bf Other prediction:} $m_H=118$ \\[2mm]
$\bullet  \qq m_H=218$ GeV \\
{\bf Authors:} Elias, Mann, McKeon \& Steele (2003)\\
{\bf Idea:} Absence of tree-level scalar-field masses \\
{\bf Techniques:} Coleman-Weinberg potential and summation of leading logarithms \\[2mm]
$\bullet  \qq m_H=226\  \pm$ 50 GeV\\
{\bf Authors:} Aranda, D\'\i az-Cruz \& Rosado (2005)\\
{\bf Idea:} unification of the weak gauge couplings at  intermediate energy $\Lambda$ and linear or quadratic relation of these to the Higgs self-coupling\\
{\bf Techniques:} renormalisation group flow up to energies of  $\Lambda\sim 10^{13}$ GeV
\\[2mm]
$\bullet  \qq m_H=241.2\  \pm$ 1.6 GeV\\
{\bf Author:} Stephan (2007)\\
{\bf Idea:} extension of Chamseddine and Connes' spectral action to $SU(4)\times SU(3)\times SU(2)\times U(1)$\\
{\bf Techniques:} operator algebras, heat-kernel expansion and renormalisation group flow up to  $\Lambda = 2\cdot 10^{4}$ GeV. The computation of the Higgs mass depends on the top mass taken to be $m_t=170.9\pm 2.6$ GeV.\\
{\bf Other predictions:} confined $SU(4)$ singlets in the TeV range, conceptual uncertainty in proper time measurements of $\Delta \tau\sim\hbar
/\Lambda= 3.3\cdot 10^{-29}$ s\\
{\bf Postdictions:} $m_W^2/(\cos^2\theta_w\,m_z^2)=1$, gluons must be massless and must have pure vector-couplings.
\\[2mm]
$\bullet  \qq m_H=250\  \pm$ 50 GeV\\
{\bf Authors:} Barbieri, Hall, Nomura \& Rychkov (2006)\\
{\bf Idea:} extending the minimal supersymmetric extension of the standard model by adding a chiral singlet with a superpotential interaction with the Higgs doublets\\
{\bf Techniques:}  renormalisation group flow up to energies of  $\Lambda\sim 10$ TeV
\\[2mm]
$\bullet  \qq m_H\approx 250 $ GeV\\
{\bf Authors:} Ne'eman \& Thierry Mieg (1982)\\
{\bf Idea:} embedding of the electro-weak Lie algebra $su(2)\op u(1)$ in the superalgebra $su(2|1)$\\
{\bf Techniques:} classical field theory \\
{\bf Postdictions:} $\sin^2\theta_w=1/4$
\\[2mm]
$\bullet  \qq m_H= 253\ \pm $ 10 GeV\\
{\bf Authors:} Arbuzov \& Zaitsev (2011)\\
{\bf Idea:} Higgs as bound state of heavy quarks\\
{\bf Techniques:} Bogoliubov compensation principle \\
{\bf Other predictions:} $m_H= 306\ \pm $ 16 GeV
\\[2mm]
$\bullet  \qq m_H=255\  \pm$ 145 GeV \\
{\bf Author:} Mahbubani (2004)\\
{\bf Idea:} split supersymmetry \\
{\bf Techniques:} fine tuning and renormalisation group flow up to energies of  $ 10^{16}$ GeV \\[2mm]
$\bullet  \qq m_H=271\  \pm$ 5 GeV\\
{\bf Authors:} Connes \& Lott (1991)\\
{\bf Idea:} derivation of the Higgs sector of the standard model from the Yang-Mills sector by generalising Euclidean  to noncommutative geometry\\
{\bf Techniques:} operator algebras. The computation of the Higgs mass depends on the top mass taken here to be $m_t=170.9\pm 2.5$ GeV.\\
{\bf Postdictions:} $m_W^2/(\cos^2\theta_w\,m_z^2)=1$, gluons must be massless and must have pure vector-couplings, $m_t>139.3$ GeV, $\sin^2\theta_w<0.543$.\\[2mm]
$\bullet  \qq m_H=275\  \pm$ 25 GeV \\
{\bf Authors:} Arbuzov, Glinka, Lednicky \& Pervushin (2007), version 1\\
{\bf Idea:} condensates, conformal cosmology \\
{\bf Techniques:} Coleman-Weinberg potential \\
{\bf Other predictions:} $m_H=124\pm 10$ GeV, version 6\\[2mm]
$\bullet  \qq m_H= 306\ \pm $ 16 GeV\\
{\bf Authors:} Arbuzov \& Zaitsev (2011)\\
{\bf Idea:} Higgs as bound state of heavy quarks\\
{\bf Techniques:} Bogoliubov compensation principle \\
{\bf Other predictions:} $m_H= 253\ \pm $ 10 GeV
\\[2mm]
$\bullet  \qq m_H=308.6\  \pm$ 0.3 GeV \\
{\bf Authors:} L\'opez Castro \& Pestieau (1995)\\
{\bf Idea:} absence of quadratic and logarithmic  divergences  in the top mass\\
{\bf Techniques:} 1-loop quantum corrections\\
{\bf Other predictions:} $m_t=170.5\pm 0.3$ GeV
 \\[2mm]
$\bullet  \qq m_H=309\  \pm$ 6 GeV \\
{\bf Authors:} Decker \& Pestieau (1979), also Veltman (1981)\\
{\bf Idea:} absence of quadratic 1-loop divergences \\
{\bf Techniques:} dimensional reduction. The computation of the Higgs mass depends on the top mass taken here to be $m_t=170.9\pm 2.5$ GeV. \\[2mm]
$\bullet  \qq m_H=317\  \pm$ 80 GeV \\
{\bf Authors:} Bazzocchi \& Fabbrichesi (2004)\\
{\bf Idea:} Flavour symmetry broken together with electro-weak symmetry, little Higgs \\
{\bf Techniques:} Coleman-Weinberg effective potential\\
{\bf Other predictions:} Many new particles including a charged scalar with mass $560\pm192$ GeV
\\[2mm]
$\bullet  \qq m_H=348.2$  GeV \\
{\bf Authors:} Bednyakov, Giokaris \& Bednyakov (2007)
\\
{\bf Idea:} $m_H=2m_t$  \\
{\bf Techniques:} arithmetic \\[2mm]
$\bullet  \qq m_H=374\  \pm$ 6 GeV \\
{\bf Author:} Xiao-Gang He (2002)\\
{\bf Idea:} no dependence of total vacuum energy (Casimir plus minimum of Higgs potential) on renormalisation scale  \\
{\bf Techniques:} Casimir effect and quantum corrections. The computation of the Higgs mass depends on the top mass taken here to be $m_t=170.9\pm 2.5$ GeV. \\[2mm]
$\bullet  \qq m_H=426$  GeV\\
{\bf Author:} Fairlie (1979)\\
{\bf Idea:} embedding of the electro-weak Lie algebra $su(2)\op u(1)$ in the superalgebra $su(2|1)$\\
{\bf Techniques:} classical field theory \\
{\bf Postdictions:} $\sin^2\theta_w=1/4$
\\[2mm]
$\bullet  \qq m_H=500\  \pm$ 100 GeV\\
{\bf Authors:} Barbieri, Hall \& Rychkov (2006)\\
{\bf Idea:} adding an inert isospin doublet of pseudo scalars\\
{\bf Techniques:} renormalisation group flow up to energies of  $\Lambda\sim 1.5$ TeV\\
{\bf Other predictions:} pseudo scalars with masses between 60 GeV and 1 TeV leading in particular to an increased width of the ordinary Higgs scalar
\\[2mm]
$\bullet  \qq m_H=515\  \pm$ 64 GeV \\
{\bf Authors:} Langguth, Montvay \& Weisz (1986)\\
{\bf Idea:} lattice gauge theory and triviality of the continuum limit \\
{\bf Techniques:} Monte Carlo simulations on $12^4$ lattices \\[2mm]
$\bullet  \qq m_H=536\  \pm$ 10 GeV \\
{\bf Authors:} Babic, Guberina, Horvat \& Stefancic (2001)\\
{\bf Idea:} no dependence of cosmological constant on renormalisation scale  \\
{\bf Techniques:}  quantum corrections. The computation of the Higgs mass depends on the top mass taken here to be $m_t=170.9\pm 2.5$ GeV. \\[2mm]
$\bullet  \qq m_H=760\  \pm$ 21 GeV \\
{\bf Authors:} Cea, Consoli \& Cosmai (2003)\\
{\bf Idea:} lattice gauge theory and triviality of the continuum limit \\
{\bf Techniques:} extrapolation from Ising limit
\\[2mm]
$\bullet  \qq m_H=1900\  \pm$ 100 GeV \\
{\bf Authors:} Iba\~nez-Meier \& Stevenson (1992)\\
{\bf Idea:} vanishing bare Higgs mass and 1-loop effective potential \\
{\bf Techniques:} autonomous renormalisation 
\\[2mm]
$\bullet  \qq m_H=10^{18}$ GeV \\
{\bf Authors:} Batakis \& Kehagias (1991)\\
{\bf Idea:} Higgs field as massive excitation of the vacuum configuration of a sigma field coupled to gravity \\
{\bf Techniques:} non-linear sigma models

\section{Upper bounds}

$\bullet  \qq m_H<123$ GeV\\
{\bf Authors:} Belyaev, Dar, Gogoladze, Mustafayev \& Shafi (2007)\\
{\bf Idea:} constrained minimal supersymmetric standard model combined with supersymmetry constraints from colliders and low energy physics and constraints on dark matter\\
{\bf Techniques:} renormalisation group equation. The top mass is taken to be $m_t=171.4 $ GeV. \\
{\bf Other predictions:} many supersymmetric particles\\[2mm]
 $\bullet  \qq m_H<125$ GeV\\
{\bf Authors:} Froggatt, Nevzorov, Nielsen \& Thompson (2008)\\
{\bf Idea:} non-supersymmetric extension of the standard model with two Higgs doublets and the multiple point principle\\
{\bf Techniques:} renormalisation group flow up to energies of  $\Lambda\sim 10^{10}$ GeV. The computation of the Higgs mass depends on the top mass taken to be $m_t=171.4\pm 2.1$ GeV.
 \\[2mm]
$\bullet  \qq m_H<125$ GeV\\
{\bf Authors:} Babu, Gogoladze, Rehman \& Shafi (2008)
\\
{\bf Idea:} minimal supersymmetric extension of the standard model plus complete vectorlike multiplets of grand unified groups\\
{\bf Techniques:} some fine tuning and renormalisation group flow up to energies of  $ 10^{16}$ GeV. The top mass is taken to be $m_t=172.6\pm1.4 $ GeV.  \\
{\bf Other predictions:} many supersymmetric particles\\[2mm]
$\bullet  \qq m_H<126$ GeV\\
{\bf Authors:} De Simone, Hertzberg \& Wilczek (2008)\\
{\bf Idea:}  Let the Higgs be the inflaton by adding a strong, non-minimal coupling $\varphi ^2R$ of the scalar to gravity. 
\\
{\bf Techniques:} effective action to 2-loop with $m_t=171.2$ GeV and its confrontation with the observed spectral index and tensor-to-scalar ratio of the Cosmic Microwave Background  \\[2mm]
$\bullet  \qq m_H<127$ GeV\\
{\bf Authors:} Carena, Nardini, Quiros \& Wagner (2008)\\
{\bf Idea:} minimal supersymmetric extension of the standard model with a light stop and electro-weak baryo-genesis \\
{\bf Techniques:} renormalisation group flow up to energies of  $\Lambda\sim 10^{16}$ GeV \\
{\bf Other predictions:} the light stop with mass $<120$ GeV\\[2mm]
$\bullet  \qq m_H<130$ GeV\\
{\bf Authors:} Okada, Yamaguchi \& Yanagida (1991)\\
{\bf Idea:} minimal supersymmetric extension of the standard model \\
{\bf Techniques:} soft supersymmetry breaking at 1 TeV and quantum corrections \\
{\bf Other predictions:} many supersymmetric particles
\\[2mm]
$\bullet  \qq m_H<130$ GeV\\
{\bf Authors:} Bento, Bertolami \& Rosenfeld (2001)\\
{\bf Idea:} introduction of a stable gauge singlet scalar of mass around 1 GeV coupled to the Higgs,  playing the role of cold dark matter and solving problems with small scale structure formation \\
{\bf Techniques:} phenomenological constraints from cosmology and particle physics \\
{\bf Other predictions:} Higgs decay into a pair of these stable light scalars
 \\[2mm]
$\bullet  \qq m_H<130$ GeV\\
{\bf Authors:} Birkedal, Chacko \& Gaillard (2004)\\
{\bf Idea:} supersymmetry and the Higgs as a pseudo-Goldstone boson of some extra global symmetry\\
{\bf Techniques:} $SU(6)$ grand unification \\
{\bf Other predictions:} many supersymmetric particles\\[2mm]
$\bullet  \qq m_H<130$ GeV\\
{\bf Authors:} Passera, Marciano \& Sirlin (2008)\\
{\bf Idea:} hypothetical errors in
the determination of the hadronic leading-order contribution to cure the present discrepancy
between experiment and prediction of the muon $g-2$  \\
{\bf Techniques:} quantum corrections by Higgs loops to precision electro-weak data with $m_t=172.6\ \pm \ 1.4 $ GeV\\
[2mm]
$\bullet  \qq m_H<130$ GeV\\
{\bf Authors:}   Nakayama, Yokozaki \& Yonekura (2011)\\
{\bf Idea:}  minimal supersymmetric extension of the standard model plus a scalar singlet \\
{\bf Techniques:}  quantum corrections \\
{\bf Other predictions:} many supersymmetric particles
\\[2mm]
$\bullet  \qq m_H<139$ GeV\\
{\bf Authors:}   Kaeding \& Nandi (1999)\\
{\bf Idea:}  a non-minimal supersymmetric extension of the standard model \\
{\bf Techniques:} gauge mediated supersymmetry breaking and quantum corrections \\
{\bf Other predictions:} many supersymmetric particles
\\[2mm]
$\bullet  \qq m_H<144$ GeV\\
{\bf Authors:}   Suematsu \& Zoupanos (2001)\\
{\bf Idea:}  a non-minimal supersymmetric extension of the standard model \\
{\bf Techniques:} non-universal soft supersymmetry breaking and quantum corrections \\
{\bf Other predictions:} many supersymmetric particles
\\[2mm]
$\bullet  \qq m_H<144$ GeV\\
{\bf Author:}   Ma (2011)\\
{\bf Idea:}  minimal supersymmetric extension of the standard model plus another $U(1)$ gauge boson\\
{\bf Techniques:}  supersymmetry breaking and quantum corrections \\
{\bf Other predictions:} many supersymmetric particles
\\[2mm]
$\bullet  \qq m_H<146$ GeV\\
{\bf Authors:}   Huo, Li, Nanopoulos \& Tong (2011)\\
{\bf Idea:}  supersymmetry \\
{\bf Techniques:} F-theory and flipped $SU(5)\times U(1)$ \\
{\bf Other predictions:} many supersymmetric particles
\\[2mm]
$\bullet  \qq m_H<150$ GeV\\
{\bf Authors:} Maloney, Pierce \& Wacker (2004)\\
{\bf Idea:} supersymmetric extension of the standard model with non-decoupling D-terms \\
{\bf Techniques:} soft supersymmetry breaking
and renormalisation group flow up to energies of  $\Lambda\sim 10^{16}$ GeV \\
{\bf Other predictions:} many supersymmetric particles and new gauge bosons with masses in the TeV range
\\[2mm]
$\bullet  \qq m_H<150$ GeV\\
{\bf Authors:} Moroi \& Okada (1992)\\
{\bf Idea:} supersymmetric extension of the standard model plus a gauge singlet \\
{\bf Techniques:} soft supersymmetry breaking
and renormalisation group flow up to energies of  $\Lambda\sim 10^{16}$ GeV \\
{\bf Other predictions:} many supersymmetric particles
\\[2mm]
$\bullet  \qq m_H<150$ GeV\\
{\bf Authors:} Carena, Nardini, Quiros \& Wagner (2008)\\
{\bf Idea:} minimal supersymmetric extension of the standard model with a light stop \\
{\bf Techniques:} renormalisation group flow up to energies of  $\Lambda\sim 10^{16}$ GeV \\
{\bf Other predictions:} the light stop
\\[2mm]
$\bullet  \qq m_H<163$ GeV \\
{\bf Authors:} Chishtie, Hanif, Jia, Mann, McKeon, Sherry \& Steele (2006)\\
{\bf Idea:} Absence of tree-level scalar-field masses \\
{\bf Techniques:} Coleman-Weinberg potential and summation of (next to$)^4$ leading logarithms \\[2mm]
$\bullet  \qq m_H<165$ GeV\\
{\bf Authors:} Delgado \& Quiros (2000)\\
{\bf Idea:} supersymmetric extension of the standard model plus one extra dimension compactified on an orbifold \\
{\bf Techniques:}  renormalisation group flow with all Higgses in the bulk  \\
{\bf Other predictions:} many supersymmetric particles
\\[2mm]
$\bullet  \qq m_H<180$ GeV\\
{\bf Authors:} Moroi \& Okada (1992)\\
{\bf Idea:} supersymmetric extension of the standard model plus extra matter multiplets \\
{\bf Techniques:} soft supersymmetry breaking
and renormalisation group flow up to energies of  $\Lambda\sim 10^{16}$ GeV \\
{\bf Other predictions:} many supersymmetric particles
\\[2mm]
$\bullet  \qq m_H<200$ GeV\\
{\bf Authors:} Espinosa \& Quiros (1998)\\
{\bf Idea:} supersymmetric extension of the standard model plus extra matter multiplets \\
{\bf Techniques:} soft supersymmetry breaking
and renormalisation group flow up to energies of  $\Lambda\sim 10^{16}$ GeV \\
{\bf Other predictions:} many supersymmetric particles
\\[2mm]
$\bullet  \qq m_H<200$ GeV\\
{\bf Authors:} Elsayed, Khalil \& Morettis (2011)\\
{\bf Idea:} supersymmetric extension of the standard model plus an extra $U(1)$ gauge boson with $B-L$ couplings and inverse seesaw \\
{\bf Techniques:}  radiative corrections \\
{\bf Other predictions:} many supersymmetric particles
\\[2mm]
$\bullet  \qq m_H<230$ GeV\\
{\bf Authors:} Bhattacharyya, Majee \& Raychaudhuri (2007)\\
{\bf Idea:} supersymmetric extension of the standard model plus one extra dimension \\
{\bf Techniques:} Kaluza-Klein and radiative corrections 
\\[2mm]
$\bullet  \qq m_H<235$ GeV\\
{\bf Authors:} Bhattacharyya, Majee \& Ray
 (2008)\\
{\bf Idea:} supersymmetric extension of the standard model plus one extra dimension, Higgs confined to the TeV brane \\
{\bf Techniques:} Kaluza-Klein and radiative corrections 
\\[2mm]
$\bullet  \qq m_H<260$ GeV\\
{\bf Authors:} Batra, Delgado, Kaplan \& Tait
 (2004)\\
{\bf Idea:} supersymmetric extension of the standard model plus a gauge singlet \\
{\bf Techniques:} soft supersymmetry breaking
and renormalisation group flow up to energies of  $\Lambda\sim 10^{16}$ GeV \\
{\bf Other predictions:} many supersymmetric particles and a charged Higgs boson lighter than the neutral one
\\[2mm]
$\bullet  \qq m_H<280$ GeV\\
{\bf Authors:} Ham, Shim, Kim \& Oh (2010)    \\
{\bf Idea:} minimal supersymmetric extension of the standard model plus a few vectorlike quarks of masses in the $300-550$ GeV range\\
{\bf Techniques:} renormalisation group flow to one-loop  \\
{\bf Other predictions:} many supersymmetric particles
\\[2mm]
$\bullet  \qq m_H<300$ GeV\\
{\bf Authors:} Babu, Gogoladze \& Kolda (2004)
\\
{\bf Idea:} minimal supersymmetric extension of the standard model plus complete vectorlike multiplets of grand unified groups\\
{\bf Techniques:} renormalisation group flow up to energies of  $ 10^{16}$ GeV  \\
{\bf Other predictions:} many supersymmetric particles\\[2mm]
$\bullet  \qq m_H<350$ GeV\\
{\bf Authors:} Batra, Delgado, Kaplan \& Tait
 (2003)\\
{\bf Idea:} supersymmetric extension of the standard model plus some new gauge bosons \\
{\bf Techniques:} soft supersymmetry breaking, some fine tuning
\\
{\bf Other predictions:} many supersymmetric particles and new gauge bosons with masses in the TeV range
\\[2mm]
$\bullet  \qq m_H<400$ GeV\\
{\bf Authors:} Litsey \& M.~Sher (2009)\\
{\bf Idea:} minimal supersymmetric extension of the standard model with a fourth generation \\
{\bf Techniques:} radiative corrections \\
{\bf Other predictions:} the fourth generation at LHC energies and many supersymmetric particles
\\[2mm]
$\bullet  \qq m_H<446$ GeV\\
{\bf Authors:} Huitu, Pandita \& Puolamaki (1997)\\
{\bf Idea:} supersymmetric extension of the left-right symmetric extension of the standard model \\
{\bf Techniques:} soft supersymmetry breaking, 1-loop corrections\\[2mm]
$\bullet  \qq m_H<450$ GeV\\
{\bf Authors:} Bhattacharyya, Majee \& Raychaudhuri (2007)\\
{\bf Idea:} supersymmetric extension of the standard model plus two extra dimensions \\
{\bf Techniques:} Kaluza-Klein and radiative corrections 
\\[2mm]
$\bullet  \qq m_H<724$ GeV\\
{\bf Authors:} Langguth \& Montvay (1987)\\
{\bf Idea:} lattice gauge theory and triviality of the continuum limit \\
{\bf Techniques:} Monte Carlo simulations on $16^4$ lattices 
\\[2mm]
$\bullet  \qq m_H<800$ GeV\\
{\bf Authors:} Appelquist \& Yee (2002)\\
{\bf Idea:} Kaluza-Kein with one or two extra comactified dimensions \\
{\bf Techniques:} computation of quantum corrections induced by Kaluza-Klein particles on precision electro-weak measurements of the flavour changing process $b\rightarrow s+\gamma$ and on the anomalous magnetic moment of the muon \\
{\bf Other predictions:} new dimensions with inverse compactification radius as low as 250 GeV
\\[2mm]
$\bullet  \qq m_H<880$ GeV\\
{\bf Authors:} Jegerlehner, Kalmykov \& Veretin (2001)\\
{\bf Idea:} Quantum corrections in the relation between $\overline {\rm MS}$- and pole-masses of the $W$- and $Z$-bosons should remain perturbative.\\
{\bf Techniques:} 2-loop corrections, asymptotic expansions 
\\[2mm]
$\bullet  \qq m_H<1008$ GeV\\
{\bf Authors:} Lee, Quigg \& Thacker (1977)\\
{\bf Idea:} unitarity requirement\\
{\bf Techniques:} partial-wave amplitude of elastic boson scattering  in lowest order of perturbation to be bounded by unity
\\[2mm]
$\bullet  \qq m_H<1020$ GeV\\
{\bf Authors:} Dicus \& Mathur (1973)\\
{\bf Idea:} unitarity requirement\\
{\bf Techniques:} partial-wave amplitude of $Z_LZ_L\rightarrow Z_LZ_L$  in lowest order of perturbation to be bounded by unity
\\[2mm]
$\bullet  \qq m_H<1400$ GeV\\
{\bf Authors:} Grzadkowski \& Gunion (2007)\\
{\bf Idea:} $W^+W^-$ scattering in the Randall-Sundrum model with one extra dimension and two 3-branes should remain perturbatively unitary after inclusion of string/M-theoretic excitations.\\
{\bf Techniques:} summation of Kaluza-Klein gravitons \\

\section{Lower bounds}

$\bullet  \qq m_H>120$ GeV\\
{\bf Authors:} Bin He, Okada \& Shafi (2009)\\
{\bf Idea:}  Let the Higgs be the inflaton by adding a strong, non-minimal coupling $\varphi ^2R$ of the scalar to gravity, add type III seesaw mechanism and demand vacuum stability and perturbativity\\
{\bf Techniques:} renormalisation group flow  \\[2mm]
$\bullet  \qq m_H>230$ GeV\\
{\bf Authors:} Barvinsky, Kamenshchik \& Starobinsky (2008)\\
{\bf Idea:}  Let the Higgs be the inflaton by adding a strong, non-minimal coupling $\varphi ^2R$ of the scalar to gravity. 
\\
{\bf Techniques:} effective action and its confrontation with the observed spectral index and tensor-to-scalar ratio of the Cosmic Microwave Background  \\

\section{Final remarks}

Our list contains 96 Higgs-mass predictions. Supersymmetry is behind 26 of them with central values between 120 and 255 GeV. Compactified additional dimensions motivate ten predictions ranging from 117 to 450 GeV.  There are three superstring inspired predictions: 117, 121 and 154.4 GeV. The embedding of the electro-weak Lie algebra $su(2)\op u(1)$ in the superalgebra $su(2|1)$ produces four predictions: 130, 161, 250 and 426 GeV. Five predictions, between 124 and 317 GeV use the Coleman-Weinberg potential. One prediction, $m_H=125$ GeV uses dynamical symmetry breaking with the Higgs being a deeply bound state of two top quarks. At the same time this model predicted two years prior to the discovery to the top its mass to be $m_t=175$.
Another prediction for the Higgs mass motivated by dynamical symmetry breaking via a neutrino condensate is at 178 GeV. We have listed four predictions from Connes's noncommutative geometry: 170, 203, 241 and 271 GeV.  Lattice gauge theories lead to two predictions: 515 and 760 GeV. Eight predictions are based on the (approximate) vanishing of particular terms related to quantum corrections:
154, 155, 200, 210, 309, 374 and 536 GeV. 

We have two lower bounds for the Higgs mass and 37 upper bounds, 26 of which come from supersymmetry.

Five predictions, one upper and one lower bound come from the recent idea that inflation is driven by the Higgs scalar together with a strong non-minimal coupling to gravity. The Higgs mass is obtained from fitting the observed spectral index and tensor-to-scalar ratio of the Cosmic Microwave Background.  

The oldest entry is: $ m_H<1020$ GeV by Dicus \& Mathur (1973).

The most precise prediction is: $ m_H= 161.8033989 $ GeV by El Naschie (2005).

The highest prediction is: $m_H=10^{18}$ GeV by Batakis \& Kehagias (1991).

The highest number of predictions by a single co-author, Gogoladze, is 12.

Three intervals are still free of Higgs-mass predictions:\\
\indent $\qq\qq600-739$ GeV, $781-1800$ GeV and $2000-10^{18}$ GeV.

In this compilation we have only considered numerical post- and predictions. Today particle physicists are used to interpret experimental numbers not only in terms of numbers like coupling constants, but also in terms of groups and representations and even in terms of Lagrangians. Only few of the listed models come with constraints on groups, representations and Lagrangians. Supersymmetric models for instance need representations for supersymmetric particles and thereby may be falsified by the LHC. However supersymmetry does not constrain the gauge group nor the Lagrangian. This is different for Connes' noncommutative geometry, which  --- just as Riemannian geometry --- puts severe constraints on the admissible Lagrangians, puts constraints on gauge groups and severe constraints on representations. In particular the Higgs representation of the noncommutative standard model is not chosen but computed to be one colourless isospin doublet. This is certainly its most startling and robust prediction and may lead to its falsification if more than one physical Higgs is found as predicted by any supersymmetric standard model. 

The first version of this compilation from 2007 contained 60 references. This seventh version has 125 references. It might well be the last one.

\end{document}